\documentclass[aps,pra,twocolumn,superscriptaddress,showpacs,floatfix]{revtex4-1}

\usepackage{savesym}
\savesymbol{iint}
\savesymbol{iiint}
\savesymbol{iiiint}
\savesymbol{idotsint}
\usepackage{amsmath}
\restoresymbol{AMS}{iint}
\restoresymbol{AMS}{iiint}
\restoresymbol{AMS}{iiiint}
\restoresymbol{AMS}{idotsint}

\usepackage{graphicx}
\usepackage{epstopdf}
\graphicspath{{./Figs/}}

\usepackage{latexsym}
\usepackage{amsmath}
\usepackage{amssymb}
\usepackage{bm}
\usepackage{wasysym}

\usepackage{ifthen}
\usepackage{color}

\newcommand{\eexp}[1]{\mathrm{e}^{#1}}

\newcommand{\be}[1]{\begin{eqnarray}{\label{e#1}}} 
\newcommand{\beq}{\begin{eqnarray}}
\newcommand{\eeq}{\end{eqnarray}} 

\newcommand{\hide}[1]{}
\newcommand{\Eq}[1]{\textcolor{blue}{{Eq.}\!\!~(\ref{#1})}} 

\newcommand{\Fig}[1]{\textcolor{blue}{Fig.}\!\!~\ref{#1}}

\begin{document}

\title{Chaos, Metastability and Ergodicity in Bose-Hubbard Superfluid Circuits}

\author{Geva Arwas and Doron Cohen}
\affiliation{Department of Physics, Ben-Gurion University of the Negev, Beer-Sheva 84105, Israel}

\begin{abstract}
The hallmark of superfluidity is the appearance of metastable flow-states that carry a persistent circulating current. 
Considering Bose-Hubbard superfluid rings, we clarify the role of ``quantum chaos" in this context. 
We show that the standard Landau and Bogoliubov superfluidity criteria fail for such low-dimensional circuits. 
We also discuss the feasibility for a coherent operation of a SQUID-like setup. Finally, we address the manifestation  
of the strong many-body dynamical localization effect. 
\end{abstract}

\maketitle

\onecolumngrid

\section{Introduction}

Circuits with condensed bosons can support superflow. Such circuits, if realized \cite{NIST2,hadzibabic,baker}, will be used as QUBITs for quantum computation \cite{brand2,Amico2014,Aghamalyan15,sfr}, or as SQUIDs \cite{BoshirPRL2013} for sensing of acceleration or gravitation. We are studying the feasibility and the design considerations for such devices. The key is to develop a theory for the superfluidity in a discrete ring \cite{Amico2014,anamaria1,Hallwood06,Paraoanu,sfa}. Such theory goes beyond the traditional framework of Landau and followers, since it involves ''Quantum chaos'' considerations \cite{kolovskiPRL,KolovskyReview,sfa}. An additional aspect concerns quantum dynamical localization, which can stabilize flows-states and suppress thermalization.

\begin{figure}[b!]
(a) \hspace{3.5cm} (b) \hspace{3.5cm} (c) \hspace{3.5cm} (d) \\
\centering
\includegraphics[width=0.20\hsize,trim=0 -20 0 0]{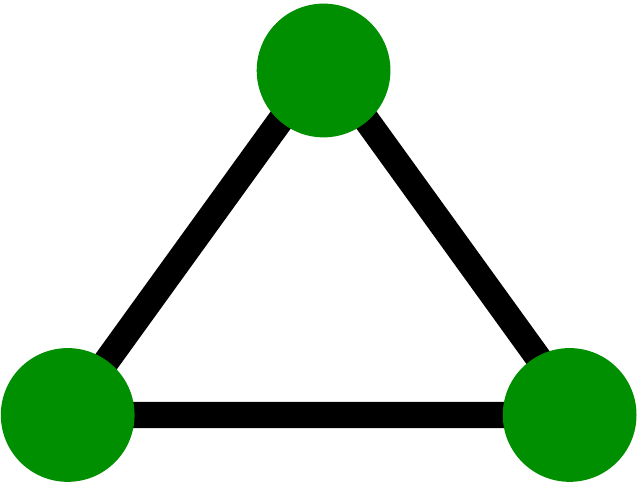} 
\hspace{6mm}
\includegraphics[width=0.20\hsize]{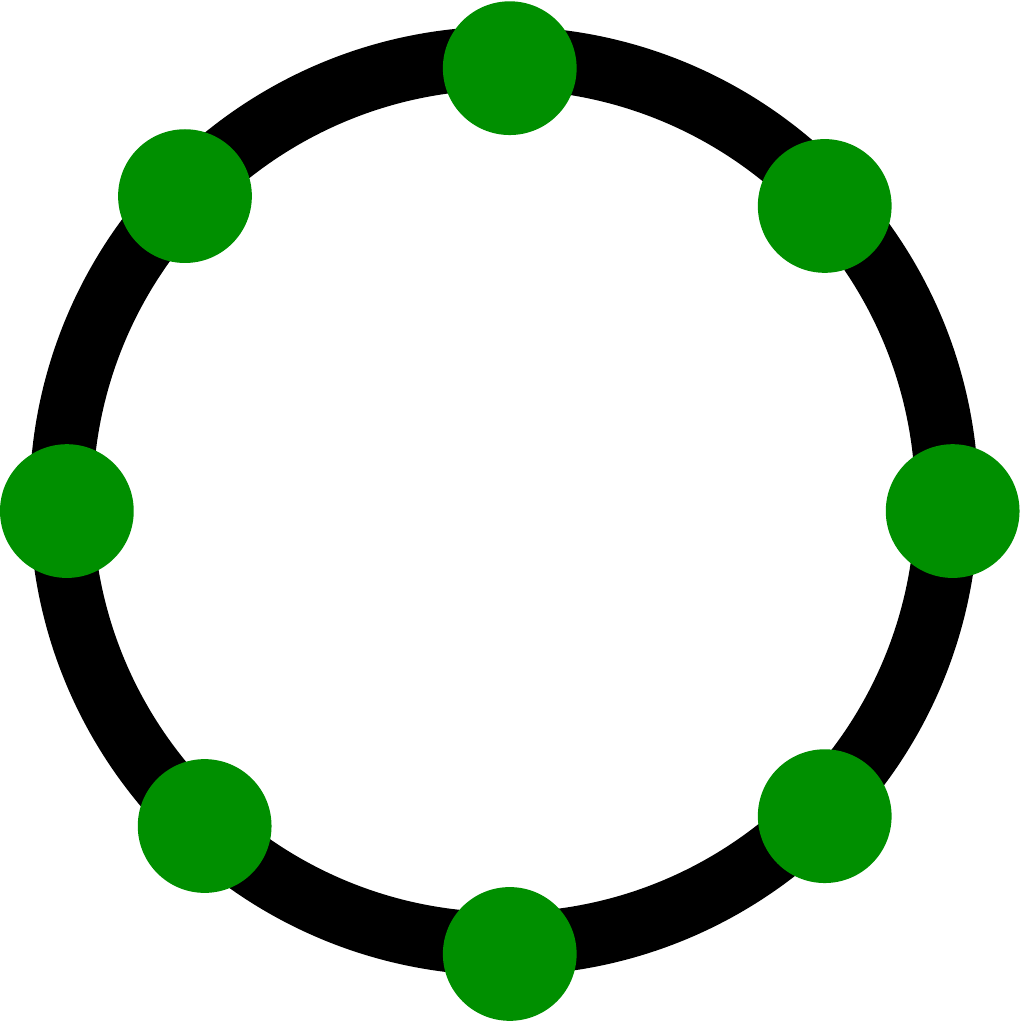}
\hspace{6mm}
\includegraphics[width=0.20\hsize]{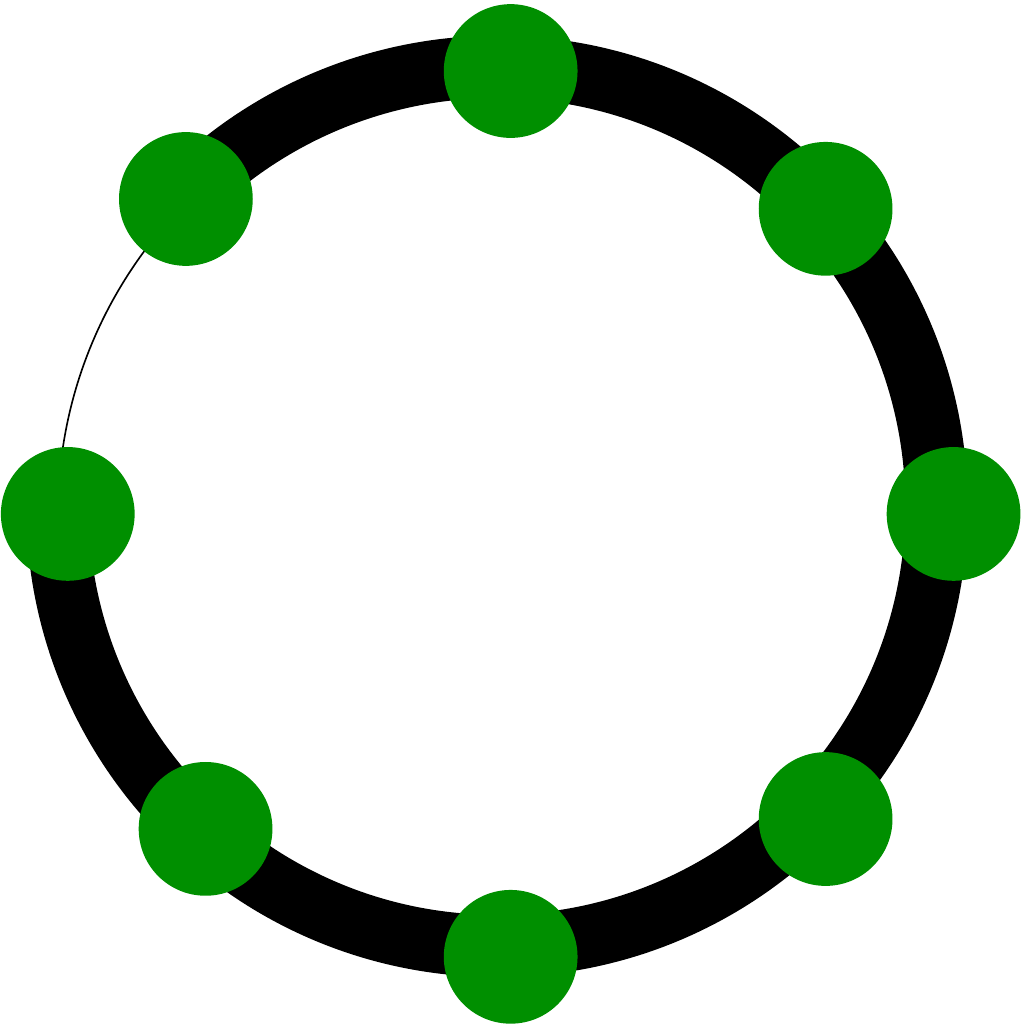}
\hspace{6mm}
\includegraphics[width=0.16\hsize]{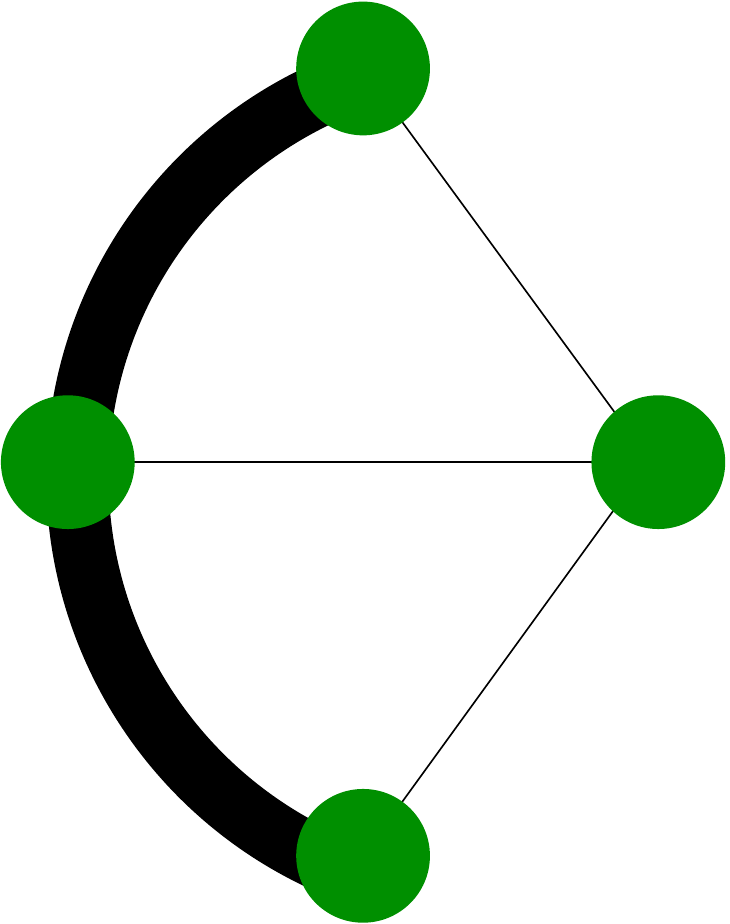}
\caption{ \label{fg1} 
{\bf Models of interest}. The dots and lines represent the Bosonic sites and the couplings. 
(a)~The $M=3$ trimer, which is the minimal model for a superfluid circuit. 
(b)~A general ${M > 3}$ ring, which exhibits high dimensional chaos and non-linear resonances. 
(c)~A SQUID-like circuit with a weak link. 
(d)~A complex composed of two weakly coupled subsystems, 
a trimer and a monomer, serve as a minimal model for thermalization.
} 
\end{figure}

In the present paper we review several results that concern Bose-Hubbard superfluid circuits \cite{sfc,sfa,sfr,mlc}. We start by introducing the model and the traditional theory for the stability of the superflow. The first configuration we consider is the smallest possible ring, with ${M=3}$ sites \cite{PhysRevE.51.2870,flach1997tunnelling,PhysRevA.63.013604,PhysRevE.67.046227,PhysRevA.73.061604,PhysRevE.86.016214,gallemi2015fragmented,sfs} \Fig{fg1}(a). We observe the existence of a novel type of superflow state, which is supported by a chaotic pond in phase-space. We then turn to discuss ${M>3}$ rings \Fig{fg1}(b), which feature high dimensional chaos and non-linear resonances. In addition we study the effect of introducing a weak link \Fig{fg1}(c). Finally we discuss the dynamics of the thermalization process, referring to \Fig{fg1}(d) as a minimal model.

\section{The model Hamiltonian}

The Bose-Hubbard Hamiltonian (BHH) is a prototype model of cold atoms in optical
lattices \cite{BHH1,BHH2}. For an $M$-site ring, 
\be{1}
\mathcal{H} \ \ = \ \ \sum_{j=1}^{M} \left[
\frac{U}{2} \bm{n}_{j}(\bm{n}_{j}-1) 
- \frac{K_j}{2} \left(\eexp{i(\Phi/M)} \bm{a}_{j{+}1}^{\dag} \bm{a}_{j} + \text{h.c.} \right)
\right]~.
\eeq
where $U$ is the on-site interaction and $j$ mod$(M)$ labels the sites of the ring. In the absence of a weak-link, we assume all the hopping frequencies are equal $K_i=K$. A weak-link means one hopping frequency is modified, say ${K_M = K' < K }$. 
The $\bm{a}_{j}$ and $\bm{a}_{j}^{\dag}$ are the Bosonic annihilation and creation operators, 
and the $\bm{n}_j \equiv \bm{a}_{j}^{\dag}\bm{a}_{j}$ are the occupation operators. The total number of particles $N= \sum \bm{n}_j $ commutes with the Hamiltonian, and is therefore conserved. 
The so-called Sagnac phase $\Phi$ appears if the ring is rotated with constant velocity \cite{fetter,NIST1}. It can be regarded as the Aharonov-Bohm flux that is associated with the Coriolis field in the rotating frame.

For the purpose of semiclassical analysis it is convenient to write the BHH 
using action-angle variables ${a_j=\sqrt{\bm{n}_j}\eexp{i\varphi_j}}$. 
For a ring with no weak link, and dropping a constant we get:
\be{2}
\mathcal{H} =  \sum_{j=1}^{M} \left[
\frac{U}{2} \bm{n}_{j}^2 
- K \sqrt{\bm{n}_{j{+}1} \bm{n}_{j}} \cos\left((\varphi_{j{+}1}{-}\varphi_{j}){-}\frac{\Phi}{M}\right)
\right]   
\eeq 
The variables $ \varphi_j$ and $\bm{n}_j$ are canonical conjugates. 
Since $N$ is a constant of motion, \Eq{e2} describes $d=M{-}1$ coupled 
degrees of freedoms (DOFs). The dimensionless parameters that characterize
the interaction are 
\be{3}
u \equiv \frac{NU}{K}, 
\hspace{3cm} 
\gamma \equiv \frac{Mu}{N^2} 
\eeq
The interaction $u$ and the flux $\Phi $ are the only dimensionless parameters which appear in the classical equations of motion. Upon quantization, the effective plank constant is ${\hbar=1/N}$, 
and the Lieb-Liniger parameter $\gamma$ is like $\hbar^2$.

The BHH in the momentum basis representation is
\be{35} 
\mathcal{H} \ = \ \sum_{k} \epsilon_k \bm{b}_k^{\dag}\bm{b}_k \ + \ \frac{U}{2M} \sum_{\langle k_1..k_4 \rangle} \bm{b}_{k_4}^{\dag}\bm{b}_{k_3}^{\dag}\bm{b}_{k_2}\bm{b}_{k_1} \ \ 
\eeq
where the $\bm{b}_k^{\dag}$ creates a particle in the $k$'th momentum orbital, 
with the energy ${\epsilon_k = -K\cos(k-(\Phi/M))}$, and the  $\langle k_1..k_4 \rangle$ summation 
is over all the $k$ values that satisfy ${k_1+k_2=k_3+k_4}$ mod$(M)$.

The hallmark of Superfluidity is the possibility to witness a 
metastable persistent current. This notion of Superfluidity does not assume a thermodynamic limit.
A coherent flow-state is created by condensing $N$~particles into 
a single momentum orbital
\beq 
| m \rangle \ \ \equiv \ \ \frac{1}{\sqrt{N ! }} \left( \bm{b}_{k_m}^{\dagger}  \right)^N | 0 \rangle 
\eeq
where $ \bm{b}_{k_m}^{\dagger} $ create a particle in a momentum orbital with winding number $m$ and wave number $k=(2\pi/M) m$. The flow states carry a macroscipically large current
\be{137}
I_{m} \ \ = \ \ \left\langle m \left|  -\frac{\partial \mathcal{H}}{\partial \Phi} \right| m  \right\rangle  \ \ =  \ \ 
N \,  \frac{K}{M} \sin \left(  \frac{1}{M} (2 \pi m - \Phi )  \right) 
\eeq
The question arises whether this current survives due to ``metastability", or decays due to ``ergodization". 
The possibility of having stable flow states (say ``clockwise" and ``anticlockwise") is the cornerstone for 
the design of a QUBIT.

\section{The traditional criteria for the stability of flow-states}

The stability of a superflow is a widely studied theme. 
The traditional approach is based on the Landau criterion \cite{landau,crit_vel1}, 
or more generally \cite{niu,smerzi,polkovnikov2005decay,dyn_stab1,dyn_stab2,dyn_stab3} on the Bogoliubov linear stability analysis.
The flow states corresponds to fixed points in phase-space: 
the $m$'th flow is situated at ${ n_1 = \cdots = n_M = N/M }$,  
meaning that the particles are distributed equally, 
and the phase differences are $\varphi_i - \varphi_{i-1} = (2\pi/M)m$.
At the vicinity of the fixed points one can linearize the classical equations 
of motion. Using the optional quantum language, adopting the Bogoliubov procedure, 
the $\bm{b}_{k_m}^{\dag}$ and $\bm{b}_{k_m}$ are replaced by $\sqrt{N}$, 
and the quadratic part is diagonalized into the form 
\be{BT}
\mathcal{H}_0  \ = \ \sum_q \omega_q \bm{c}_{q}^{\dag} \bm{c}_{q} 
\eeq
where $\bm{c}_{q}^{\dag} $ and $ \bm{c}_{q} $ 
are the Bogoliubov quasi-particles operators, given by
${\bm{b}_q^{\dagger} = u_q \bm{c}_q^{\dagger} + v_q \bm{c}_{-q}}$,  
with 
\be{4}
q=\frac{2\pi}{M} \ell, 
\ \ \ \ell=\text{integer}\neq0,  
\ \ \ -\frac{M}{2} < \ell \leq \frac{M}{2} 
\ \ \ 
\eeq
The so-called Bogoliubov frequencies are:
\be{51}
\omega_{q} = K\sin(q) \sin\left(\frac{\phi}{M}\right) + \sqrt{\left(K_q+2\frac{NU}{M}\right)K_q} \ , \ \ \ \ 
K_q \equiv 2K \sin^2\left(\frac{q}{2}\right) \cos\left(\frac{\phi}{M}\right)
\eeq
These frequencies are expressed as a function 
of the unfolded phase ${\phi = (\Phi - 2\pi m)}$.

The traditional stability criteria are based on the inspection 
of the Bogoliubov frequencies~$\omega_q$.  
Hence one can determine the stability regimes of the flow state, 
to the extent that {\em linear} stability analysis can be trusted 
(which is in fact not the case in general).  
If all $\omega_q$ have the same sign, 
the flow state are energetically stable (aka Landau stable), 
meaning that they reside in a local minima or a local maxima 
of the energy landscape. 
If one or more of the $\omega_q$ acquire an imaginary part,
the flow state become dynamically unstable, 
and one would expect a chaotic motion. 
The intermediate possibility is that all the 
Bogoliubov frequencies are real, 
but have different signs.
In such a case the dynamics is stable as far 
as the linear approximation is involved, 
but in fact this stability is endangered by 
higher order non-linear terms that have been neglected so far.

\begin{figure*}
\begin{center}
(a) \hspace{5.0cm} (b) \hspace{5.0cm} (c) \\
\includegraphics[width=0.32\hsize]{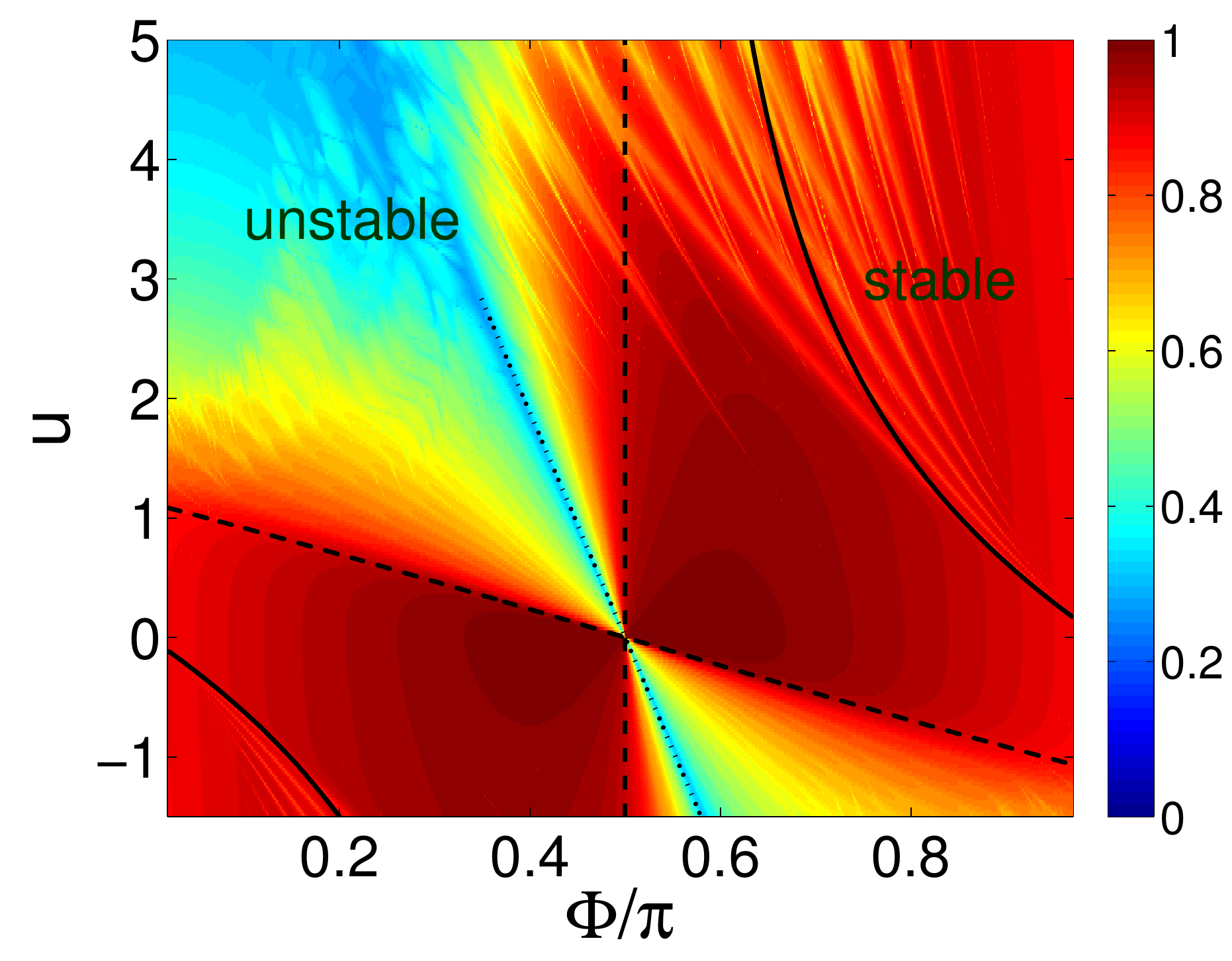}
%
\includegraphics[width=0.32\hsize]{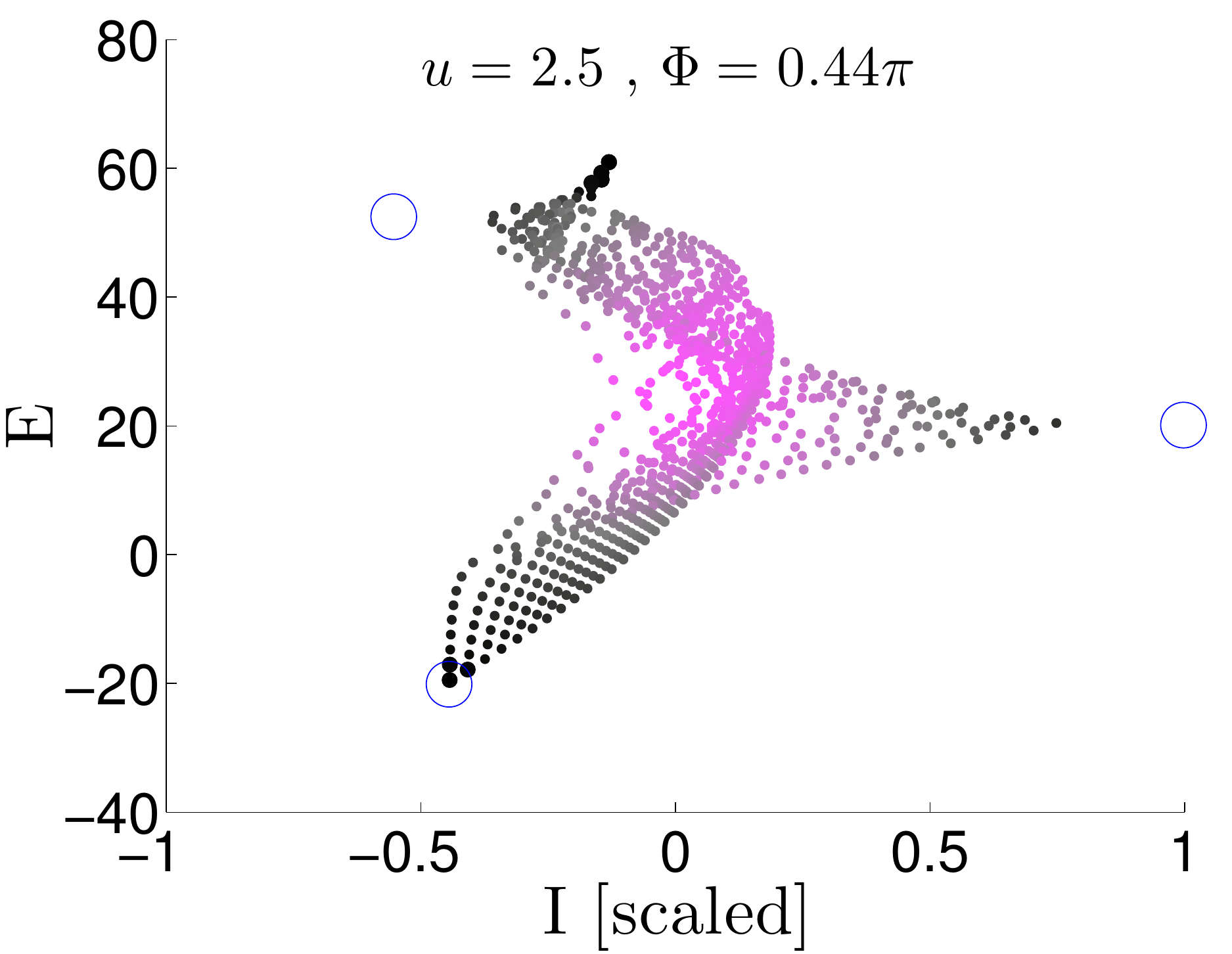}
\includegraphics[width=0.32\hsize]{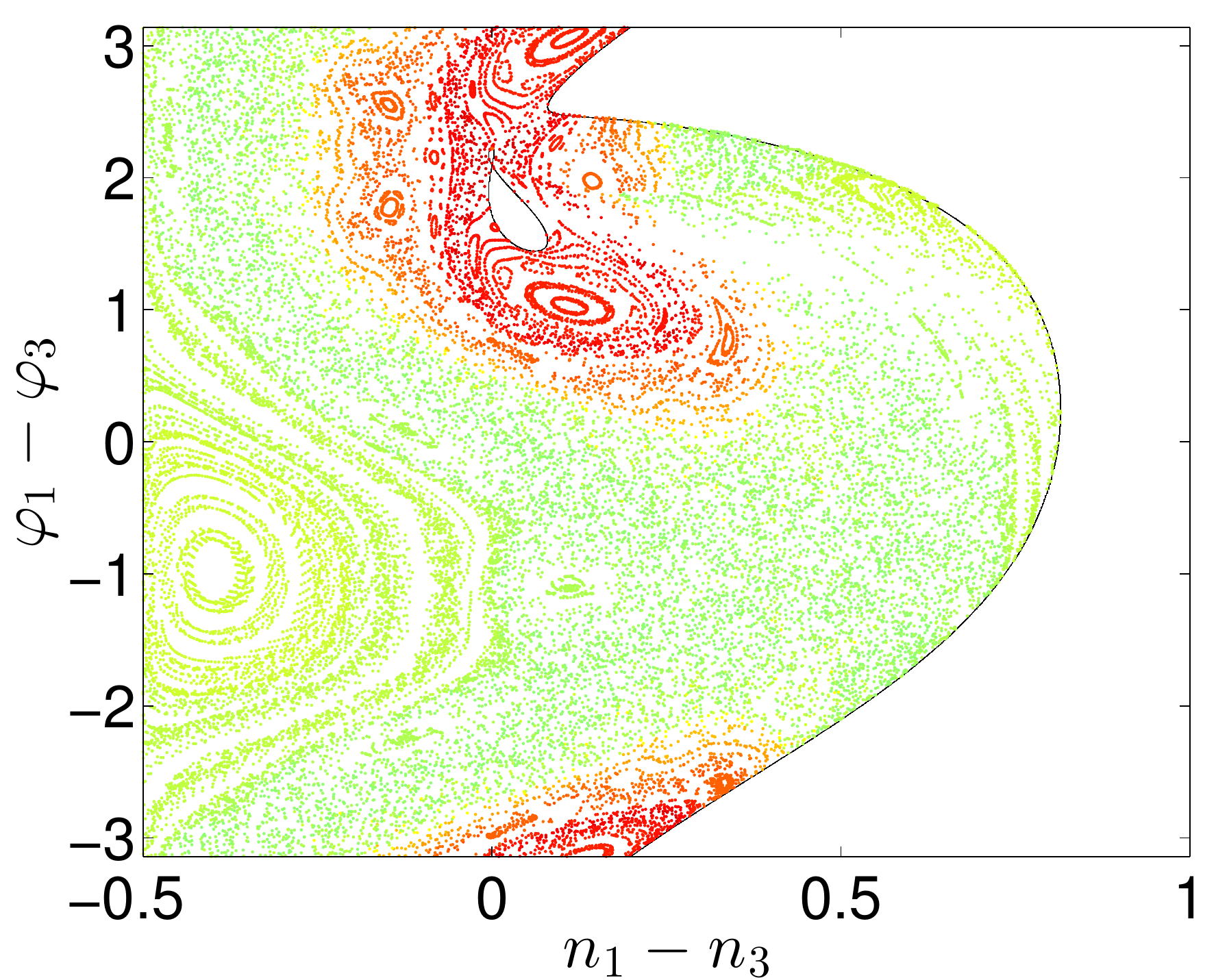}
\end{center}
\vspace*{-3.5cm} \hspace*{-12cm} $\bullet$ \vspace*{3cm}
\caption{ \label{fg2} Taken from\cite{sfc}.  
(a) Superfluidity regime diagram for $M=3$ ring with $N = 37$ particles.
The $I$ of the state that carries maximal current is imaged as a function of the model parameters $(\Phi,u)$. The solid line indicates the energetic-stability border. The dashed lines indicate the dynamical stability borders. The dotted line indicates the “swap” transition (see text). The black dot marks the $(\Phi,u)$ values used in the two other panels.
(b) Representative quantum spectrum for the $M=3$ ring with $N=42$ particles. Each point represents an eigenstate color-coded by its fragmentation (black $\mathcal{M} \sim 1 $ to purple $\mathcal{M} \sim 3 $), and positioned according to its energy and its scaled current $I/(NK/M)$. The blue circles indicate the current that would be expected by \Eq{e137}.
(c) Poincare section of $ n_3-n_2 =0 $ at the energy of the $m=1$ flow state. The flow state fixed point is located in $(n_1 - n_3 = 0  \ ; \ \varphi_1 - \varphi_3 = 2 \pi /3 )$. The solid black line marks the borders of the allowed phase-space region. The color code represents the average current for each classical trajectory.
} 
\end{figure*}

\begin{figure*}
\begin{center}
(a) \hspace{7.0cm} (b) \\
\includegraphics[width=0.4\hsize]{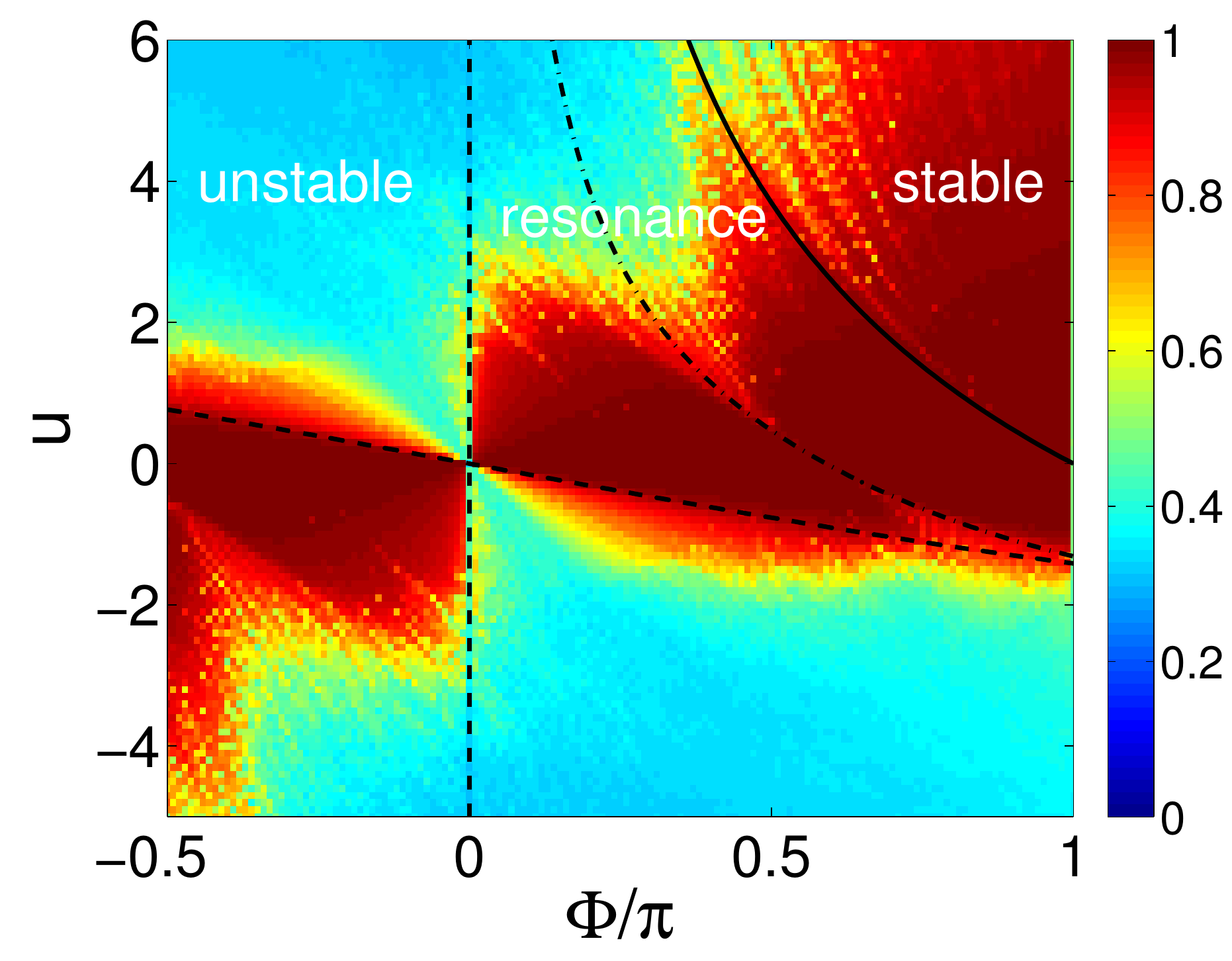}
\hspace*{1cm}
\includegraphics[width=0.4\hsize]{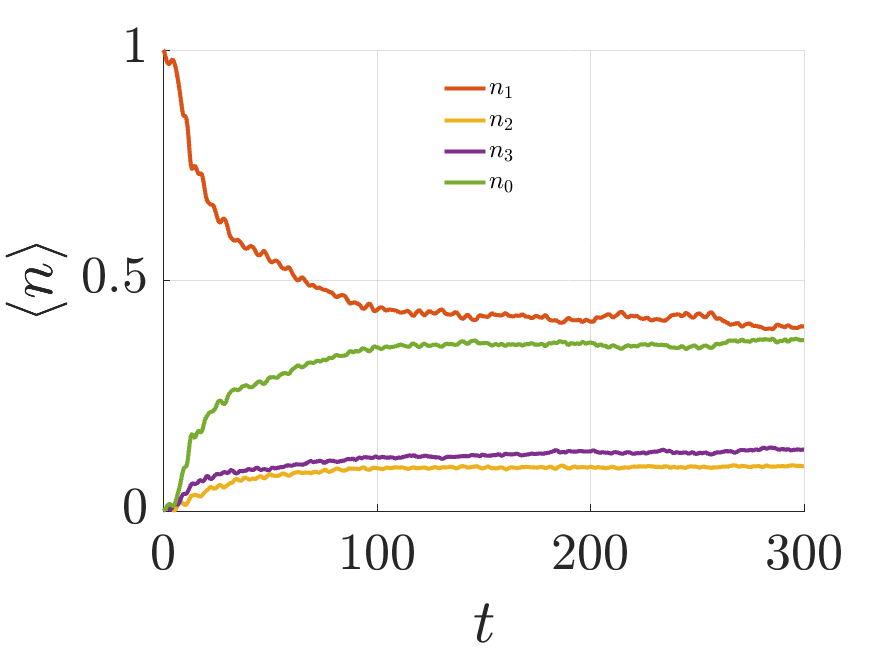}
\end{center}
\vspace*{-4.6cm} \hspace*{-8.2cm} $\bullet$ \vspace*{4.2cm}
\caption{ \label{fg3} 
(a) Superfluidity regime diagram for $M=4$ ring with $N=32$ particles. 
The long time averaged occupation $\langle  \bar{n}_1  \rangle$ of the momentum orbital, where the particles were initially condensed, is imaged as a function of the model parameters $(\Phi,u)$. The solid line indicates the energetic-stability border. The dashed lines indicate the dynamical stability borders. The resonance in \Eq{e9} plotted by a dashed-dot line. The black dot marks the $(\Phi,u)$ values of the right panel.
(b) The decay of an initially prepared $m=1$ flow state in a $M=4$ ring. Here we consider a ring with $N=120$ particles. The average occupation of the momentum orbitals are plotted as a function of time (unit are chosen such that $K=1$).  
Initially all the particles are prepared in the $n_1$ orbital. The flux $ \Phi =0.25 \pi$ and the interaction $ u \sim 2.83 $ satisfy the exact resonance condition of \Eq{e9}.
} 
\end{figure*}

Let us test the predictions of the linear stability analysis. 
In \Fig{fg2}(a) and \Fig{fg3}(a) we plot the superfluidity regime diagrams 
for ${M=3}$ ring and for ${M=4}$ ring. 
The energetic stability border is indicated by a solid line, 
while the dynamical stability border is indicted by a dashed line.
One observes that the linear stability borders 
fail to describe the color-coded numerical results: 
for the ${M=3}$ ring, dynamical \emph{instability} 
does not necessarily imply that superfluidity is diminished; 
while for the ${M=4}$ ring, 
dynamical \emph{stability} does not necessarily imply 
that superfluidity is not diminished.
These fundamental differences between rings with ${M=3}$ sites and ${M>3}$ 
sites will be explained in the next section.

\section{From KAM stability to high dimensional chaos}

The underlying classical dynamics of \Eq{e2} is chaotic.
The  $M{=}3$ ring is a $d{=}2$ system with a mixed-chaotic phase space: 
it features chaotic regions that are separated by Kolmogorov-Arnold-Moser (KAM) tori. 
This is best illustrated using a Poincare section, see \Fig{fg2}(c). 
In contrast to that, the larger ($M>3$) rings have ${d>2}$ phase-space 
with high dimensional chaos,
that features a web of non-linear resonances. In the latter case the KAM tori 
are not capable of dividing the energy shell into disjoint territories.    

Looking at the superfluidity regime diagram of the $M=3$ ring \Fig{fg2}(a) 
we see that the system has eigenstates with large current in the dynamically stable regions. But surprisingly 
we have such eigenstates also in the dynamically unstable regime. 
An example for that is given in panel~(b), where the spectrum of the many-body Hamiltonian is displayed.  
Each point represents a single eigenstate of the system: 
it is positioned according to its energy and average current, 
and color-coded by its fragmentation. 
We observe the existence of eigenstates with large current.
This is puzzling because the underlying classical motion 
in the dynamically unstable region is chaotic. 
To explain this we inspect the phase space dynamics in panel~(c), 
where we plot the Poincare section at the energy of the ${m=1}$ flow state.
The classical trajectories are color-coded by their average current, 
where red (blue) indicates large positive (negative) values. 
The section reflects the mixed phase space, featuring both chaotic and integrable regions.
We can see that the flow state fixed-point is indeed unstable, 
and a trajectory starting at its vicinity is chaotic. 
But this trajectory has large current (red). 
It does not ``ergodize" over the entire section, 
but rather confined to a small chaotic ``pond". 
This is due to the remnants of integrable structures, 
the KAM tori, which divide phase space into distinct regions, 
such that different chaotic regions are not connected.
As a result, the trajectories in the pond are chaotic, but uni-directional. 
Upon quantization, the chaotic pond can support several eigenstates
that have high current. This explains why superfluidity persists 
in the dynamically unstable region, contrary to the common expectation. 
The only region where stability is diminished in the $M{=}3$ diagram \Fig{fg2}(a) 
is along the dotted line. This line indicates a ``swap" bifurcation of separatrices \cite{sfc}.

For systems with $M>3$, meaning more then two DOF, 
it is not possible to construct a Poincare section. 
This is not merely a technical complication, but a profound difference. 
For a ${M=3}$ ring, the ${d=2}$ dimensional KAM tori divides the ${2d-1=3}$ 
dimensional energy shell into separate regions, while for ${M>3}$ this is not the case.
For example, for ${M=4}$ ring the $3$~dimensional KAM tori cannot partition the $5$~dimensional energy shell into separated regions. 
Instead, the system exhibit high-dimensional chaos, 
where all the chaotic regions are connected. 
Even if the chaos is very weak, still the stochastic regions form a connected web, 
and transport is available via Arnold diffusion \cite{LLbook,Basko2011,PhysRevLett.79.55,PhysRevLett.88.154101}.
In \Fig{fg3}(a) we plot the regime diagram for an ${M=4}$ ring. 
The main region of interest here is between the dashed and the solid lines, 
where according to the linear stability analysis the system is dynamically stable (but not energetically stable).
In principle, Arnold diffusion endangers the stability of the flow state in this entire region, 
but this is an extremely slow process. In practice, we see a significant decay 
in the dynamically stable region mainly in the vicinity of the 
dashed-dotted line, which indicates a non-linear resonance.

\section{Non-Linear resonances}

Coming back to the Bogoliubov Hamiltonian \Eq{eBT} we add the non-linear terms that have been so far ignored: 
\be{6}
\mathcal{H}\ = \ \sum_q \omega_q \bm{c}_{q}^{\dag} \bm{c}_{q} +
\frac{\sqrt{N}U}{M}  \sum_{\langle q_1,q_2 \rangle} \left[
A_{q_1,q_2} \left( \bm{c}_{-q_1-q_2} \bm{c}_{q_2} \bm{c}_{q_1}  + \text{h.c.} \right) +
B_{q_1,q_2} \left( \bm{c}_{q_1+q_2}^{\dag} \bm{c}_{q_2} \bm{c}_{q_1}   + \text{h.c.} \right)
\right]
\eeq
The summation $\langle q_1,q_2 \rangle$ excludes permutations. 
Above we have omitted 4th order terms that contain four field operators, 
because they are smaller by a factor of $\sqrt{N}$ and therefore can be neglected.
The coefficients $A$ and $B$ are functions of $(u,\Phi,M)$.
The "B" terms are the so-called Beliaev and Landau damping terms \cite{DavidsonBogoliubov,Katz2002,Iigaya2006}, while the "B" terms are usually ignored.
The former can create resonance between the Bogoliubov frequencies if 
the condition ${\omega_{q_1}+\omega_{q_2}-\omega_{q_1+q_2}=0}$ is satisfied, 
while the latter requires ${\omega_{q_1}+\omega_{q_2}+\omega_{-q_1-q_2}=0}$. 
As an example consider the $m=1$ flow state of the ${M=4}$ ring, 
for which there is a single ``$1:2$" resonance given by the $A_{q,q}$ term, where $q=2\pi/4$.
From the condition ${2\omega_q + \omega_{-2q} = 0}$ we deduce that this 
resonance appears for  ${(\Phi,u)}$ parameter values that are indicted 
in \Fig{fg2}(b) by the dashed-dot line, whose equation is 
\be{9}
u \ = \ 4 \cot \left(\frac{\Phi}{4} \right) \left[3\cos \left(\frac{\Phi}{4}\right)- \sqrt{6+2\cos \left(\frac{\Phi}{2}\right)} \right] \ \ \ 
\eeq
As implied by the color-coded numerical results, 
the width of this resonance grows as the interaction strength $u$ increases, 
and eventually covers a large fraction of the linear dynamical stability region.
In fact the width of the resonance depends on the number of particles $N$.
We have estimated \cite{sfa} that this width is proportional to $N^{-1/2}$ for fixed $u$. 
If the exact resonance condition \Eq{e9} is satisfied, 
the flow state fixed-point becomes unstable, 
and therefore an initially prepared flow state will decay, irrespective of~$N$. 
An example for the time dependence of this decay is provided by \Fig{fg3}(b).

\begin{figure*}
\begin{center}
(a) \hspace{6.0cm} (b) \\
\includegraphics[width=0.4\hsize]{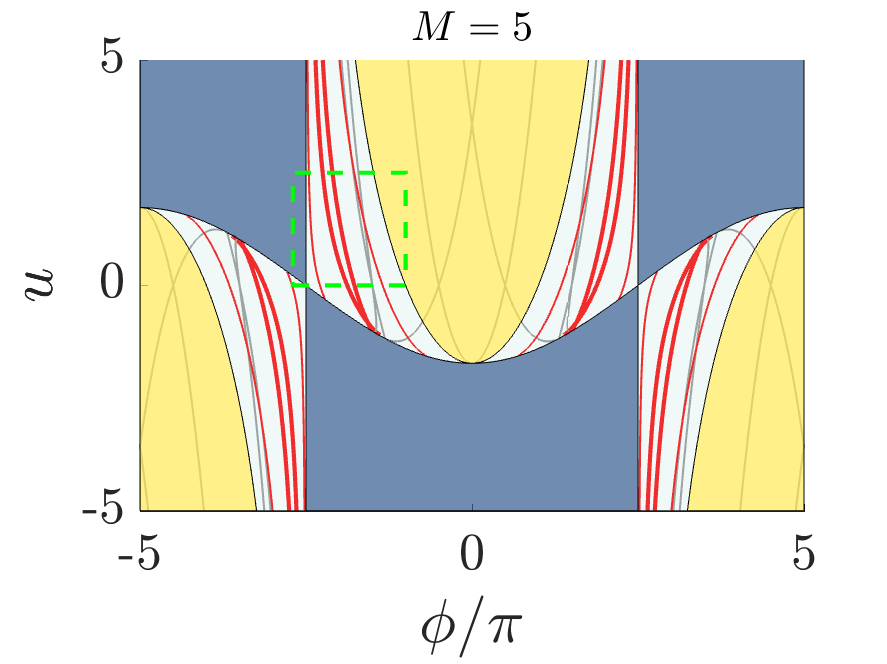}
\includegraphics[width=0.4\hsize]{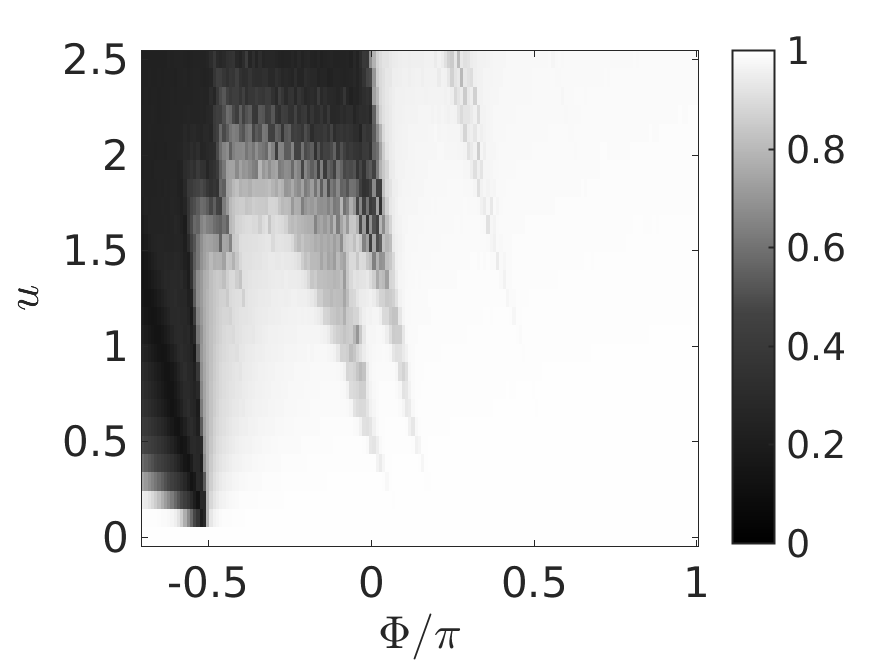}
\caption{ \label{fg4} 
(a) Taken from \cite{sfa}. The superfluidity regime diagram for $M=5$. Yellow and gray indicate the Landau-stability
and linear-instability regions. Nonlinear resonances of the "A" and "B" types are indicted by red and gray lines, respectively. The additional thin red lines are fourth-order resonances. The green
squares mark regions of interest that will be explored in the right panel.
(b) The survival of a prepared $m=1$ flow state in a $M=5$ ring with $N=50$ particles. Initially all the particles are condensed in the $n_1$ momentum orbital. The “survival” (see
text) is imaged as a function of $(\Phi,u)$.
} 
\end{center}
\end{figure*}

In a larger $M$ system we have more degrees of freedom, 
and therefore more resonances. In \Fig{fg4}(a) we image the $M=5$ regime diagram. 
The background color indicates the linear stability regimes: 
yellow indicates energetic stability, 
grey indicates dynamical instability, 
and the middle region indicates linear dynamical stability. 
The red lines are the "A" type resonances that destabilize the flow states, 
while the grey lines are the "B" resonances.

In \Fig{fg4}(b) we focus on the parametric range marked by a green rectangle in \Fig{fg4}(a), 
and plot the ``survival" of a prepared coherent flow state. 
We define the ``survival" as the normalized occupation of
the flow state orbital, as deduced from inspecting the long-time dependence. 
We can see significant decay near the two "red" resonances, 
which completely overlap for a sufficiently large $u$ values. Note that the "B" type (gray) resonances barely affect. It can be proven \cite{sfa} that they are unable to destroy the stability.


\section{Coherent Rabi oscillations}

So far we have considered the stability of flow states. In this section we ask whether two quasi-degenerate flow states can form an effective two-level system (TLS). If such a TLS is formed, we expect to observe coherent Rabi oscillations between the two macroscopically distinct flow states, and the device could possibly serve as a qubit. In particular the $m=0$ and the $m=1$ flow states are quasi-degenerate provided $\Phi=\pi$, and an effective TLS is formed at the bottom of the spectrum.
The coupling $\Delta_s$ between the two flow states typically decreases exponentially with the number of particles, hence the period of the Rabi oscillations $2\pi / \Delta_s $ becomes too large for practice applications.
One possible way to improve the control over $\Delta_s$ is by modifying one of the coupling, such as to have a weak link within the circuit, 
see text after \Eq{e1}. The semiclassical coordinates that describe the weak-link are 
the phase difference ${\bm{\varphi} = (\bm{\varphi}_M - \bm{\varphi}_1)}$, 
and and the conjugate ${\bm{n} }$ as in SQUID circuit. 

For $M \gg 1 $ one can approximate the remaining DOFs as a Caldeira-Leggett bath,  
and the Hamiltonian takes the of the Josephson Circuit Hamiltonian (JCH) 
\be{52}
\mathcal{H}_{\text{JCH}} \ \ = \ \ E_C \ \bm{n}^2 
+ \frac{1}{2}E_L \bm{\varphi}^2
- E_J \ \cos(\bm{\varphi}-\Phi)  
+ \mathcal{H}_{\text{bath}}
\eeq
with ${E_C=U}$, and $E_L=[(N/M)/(M-1)]K$, and $E_J=(N/M)K^{\prime}$. 
The condition for having at least one pair of metastable flow-states at flux $\Phi=\pi$, i.e. a double well in the energy landscape, 
is ${\alpha > 1}$ where  ${\alpha \equiv E_J/E_L =  (M-1)K^{\prime}/K}$. 
The dissipation coefficient that characterized the Caldeira-Leggett bath is 
\be{53}
\eta \ = \ \frac{\pi}{\sqrt{\gamma}}
\eeq
where $\gamma$ has been defined in \Eq{e3}. A full derivation of the JCH coeficients and the bath Hamiltonian is given in \cite{sfr} (see also \cite{Hekking,Rastelli,Amico2014}) The condition for witnessing coherent oscillations is ${\eta < \pi}$, 
which requires ${\gamma > 1}$. This is clearly problematic because it coincides with the 
border of the Mott regime, where the ring is likely to be a Mott insulator, depending of the ratio $N/M$.

\begin{figure*}
\begin{center}
(a) \hspace{5.5cm} (b)  \\
\includegraphics[width=0.35\hsize]{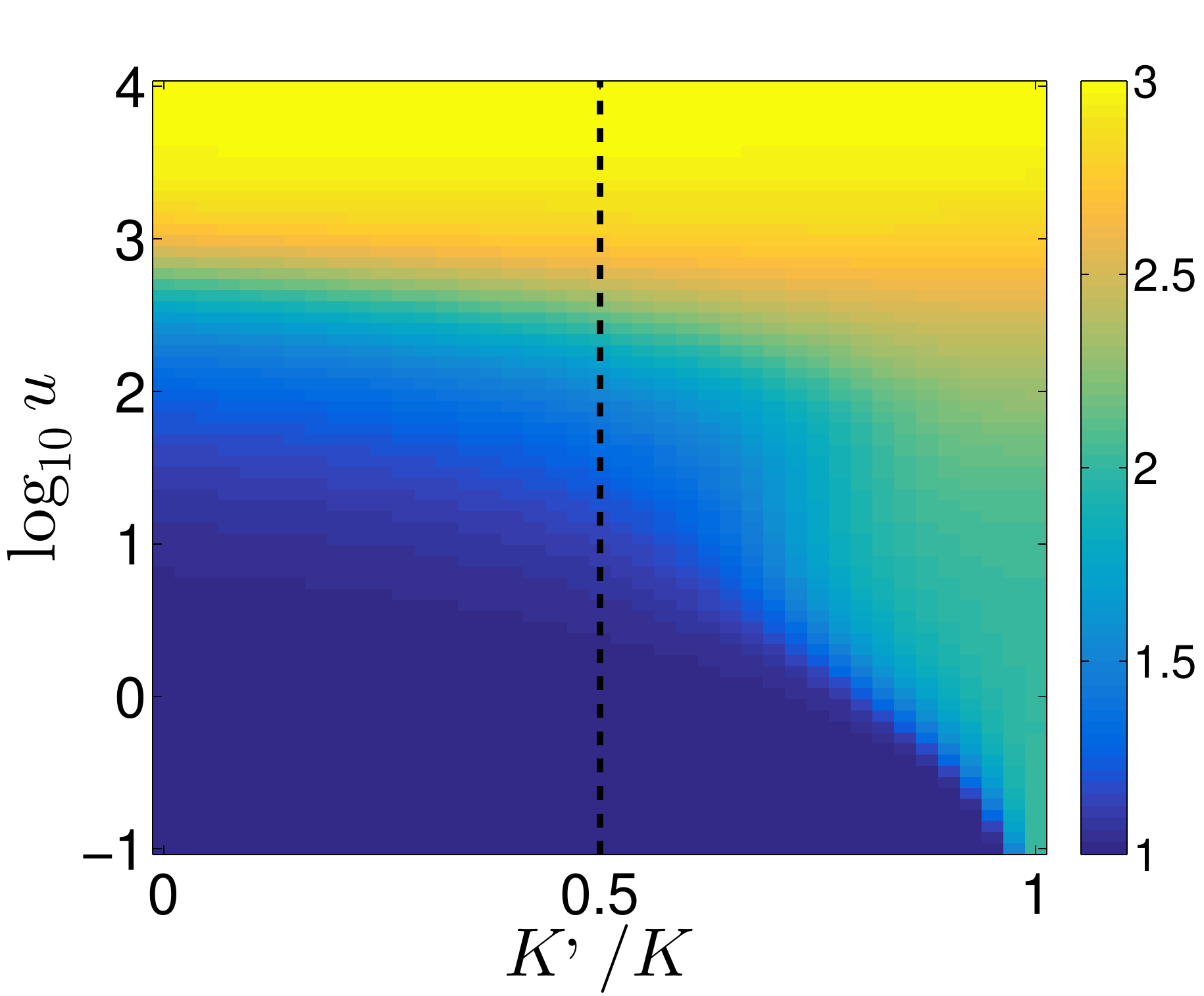}
\includegraphics[width=0.35\hsize]{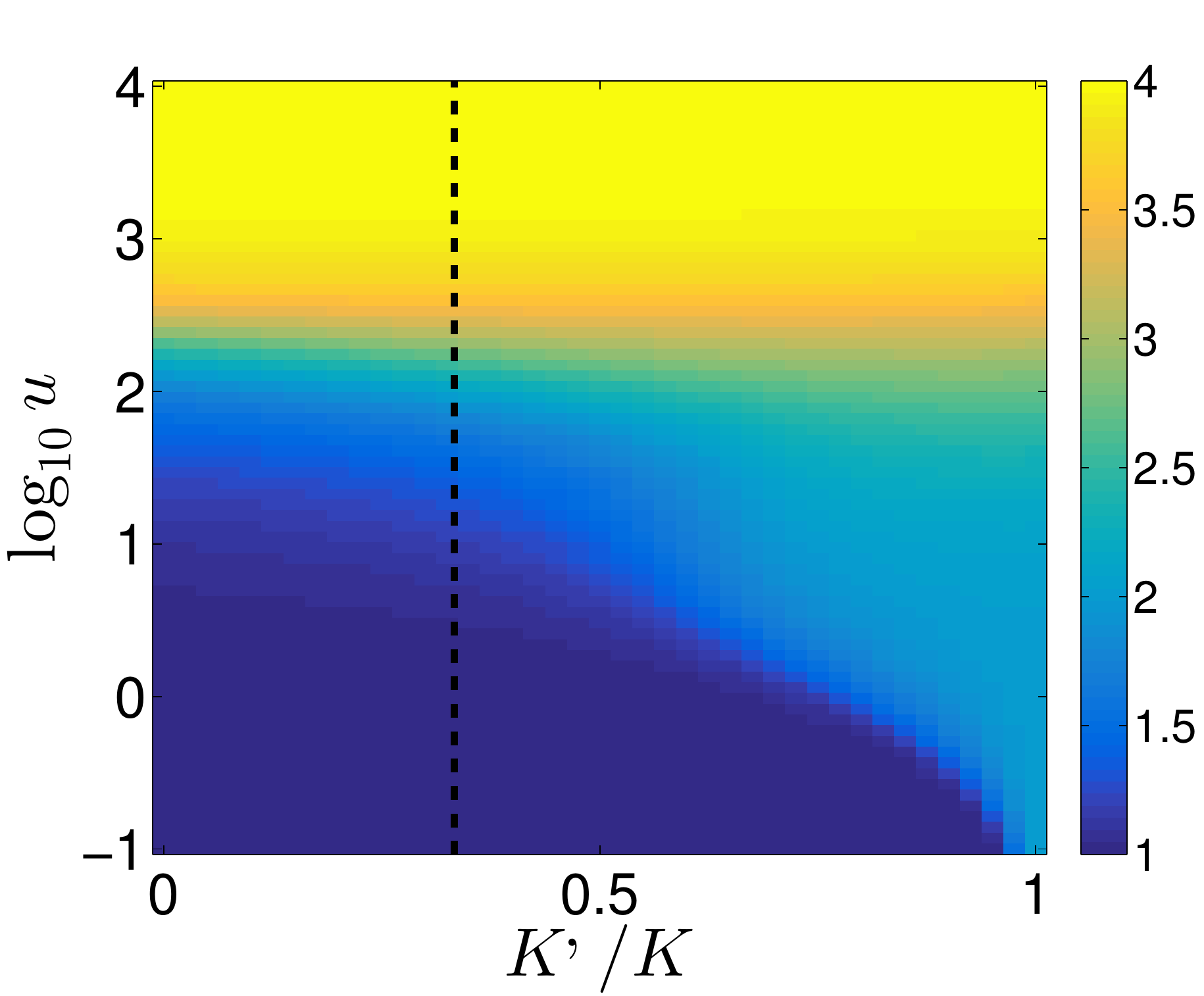}
\caption{ \label{fg5} Taken from \cite{sfr}. 
The fragmentation ($\mathcal{M}$) of the ground state is imaged as a function 
of $u$ and $K'/K$ for $M{=}3$ ring with $N{=}30$ particles (left) 
and for $M{=}4$ ring with $N{=}20$ particles (right). 
The value $\mathcal{M}=1$ indicates a coherent state (all particles are condensed 
in a single orbital). The value of ${\mathcal{M} \sim 2}$ indicates quasi degeneracy 
of the ground state (a doublet of flow-states). The value ${\mathcal{M} \sim M}$ 
indicates a fragmented state: here it is due to the quantum Mott transition.  
The vertical dashed line corresponds to the ${\alpha=1}$ border, 
which in the absence of a Mott transition would become valid for large~$u$.  
} 
\end{center}
\end{figure*}

We are therefore motivated to consider small rings where the other DOFs are not effective like a ``bath". 
What does it mean small? Clearly we want to have a ring for which \Eq{e52} is inapplicable. At this stage 
one should realize that the JCH approximation assumes that the chaos threshold (in energy) is well above 
the height of the dividing barrier; hence the dynamic in leading order is like having a single degree of freedom. 
In \cite{sfr} we have argued that this is not the case for a ring that 
has less than 6~sites. For ${M<6}$ a full phase-space analysis is required.  
In particular we have considered $M=3,4$ rings with a weak link. In order to determine the range of parameters 
for which a {\em coherent} TLS operation is feasible we have used a fragmentation-based measure.   
The fragmentation of the ground state is defined as $\mathcal{M}=[\text{trace}(\rho^2)]^{-1}$, 
where $ \rho_{ij} = \langle  \bm{a}_{i}^{\dag} \bm{a}_{j}  \rangle /N$ is the one-body reduced probability matrix.
In \Fig{fg5} we image $\mathcal{M}$ for $\Phi=\pi$. If an effective TLS is formed at the bottom of the spectrum, 
we expect the ground state to be a macroscopic superposition of two flow states, hence ${\mathcal{M}  \sim 2}$. 
If the weak link is too weak, the TLS breaks down, and the ground state is a coherent state with ${\mathcal{M} \sim 1}$. 
We see that the $\alpha$ border is slightly higher then ${\alpha=1}$, which reflects the high-dimensional 
nature of the double well in phase-space. For large $u$ we see that ${\mathcal{M} \sim M}$, indicating a maximally fragmented Mott state.

\section{Thermalization}

In the classical treatment any connected chaotic region ergodizes, 
hence it is not likely to witness dynamical metastability
for an ${M>3}$ model. Even for weak chaos we have Arnold diffusion. 
Still this Arnold diffusion is very slow and in practice possibly cannot be observed.
Furthermore, upon quantization it is likely to be completely suppressed 
due to a dynamical localization effect.

It is in fact more interesting to study the dynamical localization effect 
for the minimal model that is illustrated in \Fig{fg1}(d). 
Consider a 3-site Bose-Hubbard subsystem (trimer) with $x$ particles, 
coupled weakly to an additional site (monomer) with $N-x$ particles. 
In \cite{tmn} it has been demonstrated that the probability distribution $\rho(x)$ 
obey a Fokker-Planck equation in the classical limit; 
with an effective diffusion coefficient that requires 
a resistor-network perspective. 
However, in the quantum case, the spreading is suppressed 
due to a strong quantum localization effect
if $x$ is below or above some threshold values. 
Using a semiclassical approach it is possible to determine 
these mobility edges, and the localization volume 
in phase space \cite{mlc}.

\section{Conclusions}

We have clarified the role of ``chaos" for the metastability criteria of flow states, and for the possibility to witness Rabi oscillations in a SQUID-like setup. Additionally we considered both coherent and stochastic-like features in the dynamics of the thermalization process. Our main observations are: {\bf (1)}~Instability of flow states for a three sites ring is due to swap of separatrices; {\bf (2)}~For rings with more than three sites it has to do with a web of non-linear resonances; {\bf (3)}~It is not likely to observe coherent operation for rings that have a weak link and more than five sites;  {\bf (4)}~Strong many-body dynamical localization may enhance the stability, and suppress stochastic-like thermalization.





%


\end{document}